\documentclass[aps,prl,twocolumn]{revtex4-1}
\usepackage[T1]{fontenc}
\usepackage[latin1]{inputenc}
\usepackage{lmodern}
\expandafter\let\csname equation*\endcsname\relax
\expandafter\let\csname endequation*\endcsname\relax
\usepackage{amsmath}
\usepackage{esint}
\usepackage{amssymb}
\usepackage{graphicx}
\usepackage{xcolor}
\usepackage{natbib}
\usepackage{bm}
\usepackage{bbold}
\usepackage[mathscr]{euscript}
\usepackage[cal=boondoxo]{mathalfa}
\usepackage{comment}
\def\ii{{\rm i}}

\def\rb{{\bf r}}

\def\braket#1{\mathinner{\langle{#1}\rangle}}

\def\pM{\mathrel{\raise 2pt \hbox{\tiny(}\!\raise 1pt \hbox{+}\settowidth {\dimen03} {+}\hskip-\dimen03 \raise -2.4pt \hbox {$-$} \!\raise 2pt \hbox{\tiny)}}}

\begin{document}
\title{Flat bands and chiral optical response of moir\'e insulators}
\author{H. Ochoa}
\author{A. Asenjo-Garcia}
\affiliation{Department of Physics, Columbia University, New York, NY 10027, USA}

\date{\today}
\begin{abstract}
We present a low-energy model describing the reconstruction of the electronic spectrum in twisted bilayers of honeycomb crystals with broken sublattice symmetry. The resulting moir\'e patterns are classified into two families with different symmetry. In both cases, flat bands appear at relatively large angles, without any \textit{magic angle} condition. Transitions between them give rise to sharp resonances in the optical absorption spectrum at frequencies well below the gap of the monolayer. Owing to their chiral symmetry, twisted bilayers display circular dichroism -- different absorption of left and right circularly-polarized light. This optical activity is a nonlocal property determined by the stacking. In hexagonal boron nitride, sensitivity to the stacking leads to strikingly different circular dichroism in the two types of moir\'es. Our calculations exemplify how subtle properties of the electronic wavefunctions, encoded in current correlations between the layers, control physical observables of moir\'e materials.
\end{abstract}
\maketitle

Twisting and sliding the layers of van der Waals materials creates superstructures with emergent properties. A singular case is that of twisted bilayer graphene, i.e., two graphene layers rotated with respect to each other an angle $\theta$ \cite{portu}. In the regime of small twist angles, the relative shift in the periodicities of each individual lattice creates a moir\'e beating pattern. Interference of Dirac electrons in the associated superlattice potential reshapes the conical dispersion of the low-energy bands. As the twist approaches the \textit{magic angle} $\theta\sim 1.1^{\textrm{o}}$ \cite{numerics,MacDonald}, the reconstructed Dirac cones evolve into flat bands, favoring the emergence of electronic correlated states \cite{PJH}.

In twisted bilayers whose components have a gap (e.g., semiconducting transition-metal dichalcogenides \cite{MacDonald_TMDC,TMDC} or hexagonal boron nitride), flat bands appear without any special alignment condition due to the existence of a natural energy scale for electron confinement in the original spectrum. This is in stark contrast with the case of twisted bilayer graphene, which remains gapless after twisting, as the Dirac cones are preserved by the emergent symmetries of incommensurate superstructures \cite{emergent}. The latter are captured by the \textit{continuum model} \cite{portu,MacDonald}, which accurately describes the electronic spectrum at small twist angles.

A common property of these systems is that twisting the layers breaks all the inversion planes/centers, reducing the symmetry to a purely rotational (i.e., chiral) group. Chirality gives rise to optical activity, and is usually measured through different responses to circularly polarized light \cite{CD_exp}.  This circular dichroism (CD) results from current counterflows in response to an electric field \cite{Brey}, which engenders an in-plane magnetization density that rotates the polarization of the incident light \cite{Stauber}. 

Here, we study the optical activity of twisted bilayers with broken sublattice symmetry (i.e., where each sublattice is occupied by different atomic species). We employ a low-energy continuum model and go beyond the single band approximation~\cite{MacDonald_TMDC,Tong_etal} by including coherences across the gap, to capture optical processes. Applying our model to twisted bilayers of hexagonal boron nitride (hBN), we find significant CD at frequencies that are resonant with the energy difference between the lowest flat bands. This optical activity strongly depends on the stacking, as interlayer current correlations are mostly determined by the hopping between atoms of different polarity.

\begin{figure}
\centerline{\includegraphics[width=\linewidth]{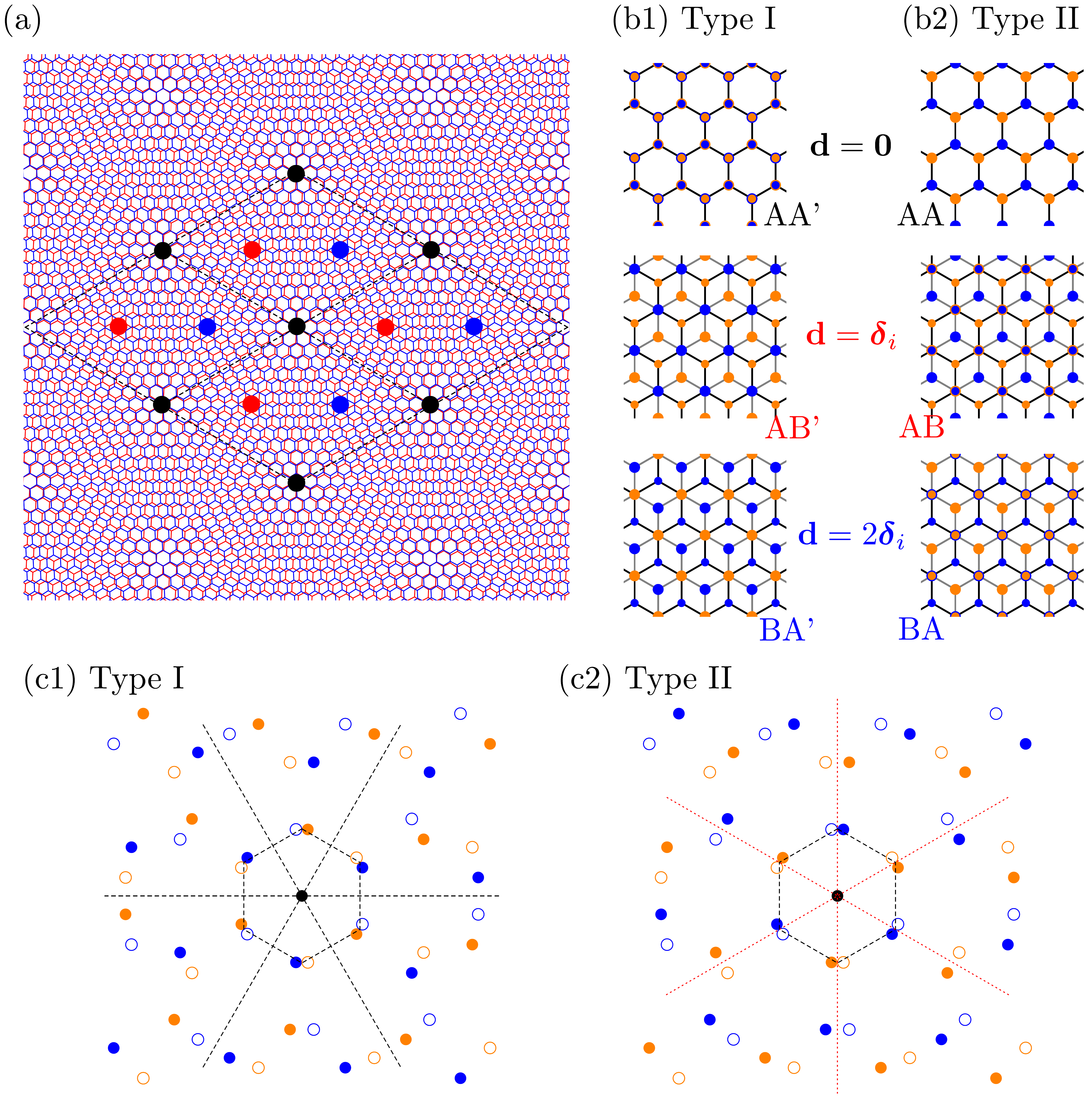}}
\caption{\textbf{Moir\'e patterns and symmetries.} (a) Moir\'e pattern resulting from a twist of $\theta=5^{\textrm{o}}$ and relative translation $\mathbf{u}=0$. Black, red and blue dots signal local high-symmetry stackings. (b) Starting from either AA' or AA configurations, a high-symmetry stacking is generated by sliding one layer with respect to the other a nearest-neighbor vector $\boldsymbol{\delta}_i$. (c) Point-group operations in both types of moir\'e structures, where open (filled) circles represent atoms on the top (bottom) layer. Dashed lines represent in-plane C$_2$ rotation axes and the black point is the C$_3$ rotation. In (b,c) orange and blue circles represent nitrogen and boron sites, respectively.}\label{fig:moire}
\end{figure}

There are two families of twisted structures with different symmetry, as shown in Fig.~\ref{fig:moire}. Any structure can be generated by a relative rotation of angle $\theta\in[-30^{\textrm{o}},30^{\textrm{o}}]$ taken along a common center of the hexagons, followed by a relative translation $\mathbf{u}$, starting from either AA' or AA stacking configurations, see Fig.~\ref{fig:moire}~(b). In the first case, ions of opposite polarity lie on top of each other before the twist; these are moir\'e structures of type I. The other case, where ions of same polarity initially sit on top of each other, are moir\'e structures of type II. The shift in periodicity of  the individual lattices define the primitive wavevectors $\mathbf{G}_i$ of the associated beating pattern, $\mathbf{G}_i=R(-\theta/2)\mathbf{g}_i-R(\theta/2)\mathbf{g}_i=2\sin(\theta/2)\mathbf{g}_i\times\hat{\mathbf{z}}$, where $\mathbf{g}_i$ are the primitive vectors of the reciprocal lattice prior to the rotation, and $R$ is the associated matrix. Within the moir\'e supercell, highlighted with dashed lines in Fig.~\ref{fig:moire}~(a), structures of type I explore the initial AA' (black dots), AB' (red dots) and BA' (blue dots) registries. In type-II structures, we have instead AA, AB and BA registries. 

\begin{figure*}
\centerline{\includegraphics[width=\linewidth]{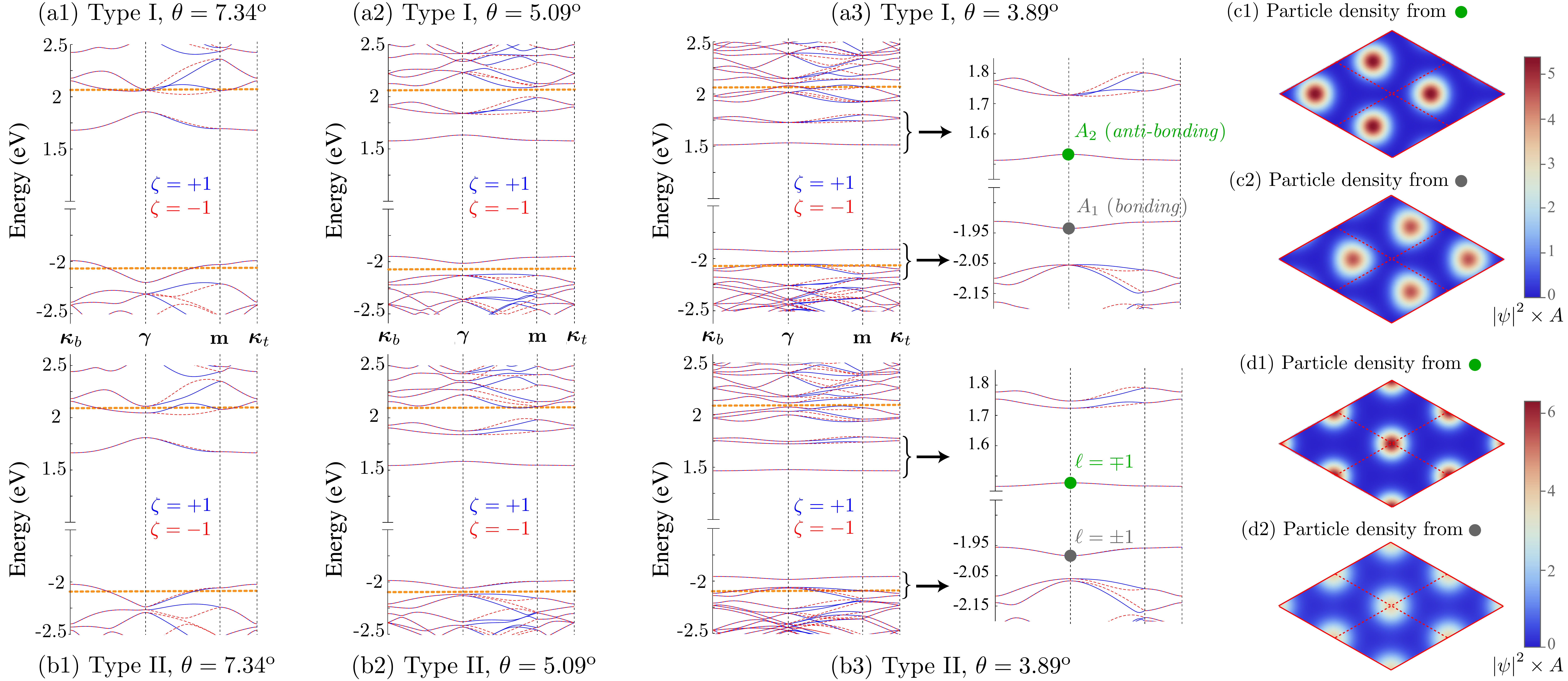}}
\caption{\textbf{Band structure and envelope wave functions of twisted hBN bilayers.} (a) Band structure of type-I moir\'es with $\theta=7.34^{\textrm{o}}$ (a1), $\theta=5.09^{\textrm{o}}$ (a2), and $\theta=3.89^{\textrm{o}}$ (a3). The number of harmonics is truncated to $n_{\mathbf{G}}=37$ (a1 and a2) and $n_{\mathbf{G}}=61$ (a3). Dashed orange lines indicate the position of the band edges in the energetically most preferable configuration, AA' and AB/BA stacking for type I and type II structures, respectively. The right panel shows a zoom of the lowest energy bands. $A_1$ and $A_2$ labels the irreducible representation of the wave functions at $\boldsymbol{\gamma}$. (b) The same for type-II moir\'es, with $n_{\mathbf{G}}=37$ (b1), $n_{\mathbf{G}}=61$ (b2), and $n_{\mathbf{G}}=91$ (b3). The angular momentum numbers $\ell$ label the $C_3$ representations of the wave functions at $\boldsymbol{\gamma}$. (c, d) Total amplitude $\sum_{\alpha,\mu} |\psi_{+,\mathbf{q=\boldsymbol{\gamma}}}^{\alpha,\mu}\left(\mathbf{r}\right)|^2$ of the wave function at $\mathbf{q}=\boldsymbol{\gamma}$ for the lowest energy bands in type-I (c) and type-II (d) structures for $\theta=3.89^{\textrm{o}}$.}\label{fig:bands}
\end{figure*}

At small twist angles, the low-energy electronic states around valley $\zeta$ can be described by a continuum Hamiltonian of the form\begin{align}
\label{eq:Hamiltonian}
   \hat{\mathcal{H}}_{\zeta}\left(\mathbf{r}\right)=\left[\begin{array}{cc}
  \hat{\mathcal{H}}_{\zeta}^{(t)}\left(\mathbf{r}\right) & \hat{T}_{\zeta}\left(\mathbf{r}\right)\\
  \hat{T}_{\zeta}^{\dagger}\left(\mathbf{r}\right) & \hat{\mathcal{H}}^{(b)}_{\zeta}\left(\mathbf{r}\right)
   \end{array}\right],
\end{align}
where blocks in the diagonal describe the top ($t$) and bottom ($b$) layers, and the off-diagonal blocks account for interlayer hopping processes. The Hamiltonian acts on a Hilbert space of envelope wave functions, which are smooth on the scale of the microscopic lattice spacing, $a$. Next, we derive the model for hBN, where the edges of the $\pi$ bands reside around the two inequivalent corners $\mathbf{K}_{\pm}$ of the hexagonal Brillouin zone. The sublattice, layer, and valley ($\zeta=\pm1$ for $\mathbf{K}_{\pm}$) are incorporated as internal quantum numbers. The blocks in the diagonal follow from the tight-binding Hamiltonian $\hat{\mathcal{H}}\left(\mathbf{k}\right)=t\,\text{Re}\{ \phi(\mathbf{k})\}\hat{\sigma}_x-t\,\text{Im}\{ \phi(\mathbf{k})\}\hat{\sigma}_y+\Delta/2\,\hat{\sigma}_z$, where $\phi\left(\mathbf{k}\right)=\sum_i e^{i\boldsymbol{\delta}_i\cdot\mathbf{k}}$, and the sum is extended to the three vectors connecting first nearest neighbors. The Pauli matrices act on sublattice space; $t\approx 2.8$ eV \cite{hopping} represents the intralayer hopping, while $\Delta$ accounts for the energy difference between the two ions in the unit cell. In type-I structures, the atomic species on each sublattice are swapped in the bottom layer, so the sign of $\Delta$ is inverted.

We now introduce the effect of the twist. The positions of the valleys are shifted to $\mathbf{K}_{\zeta}^{(t,b)}=R(\pm \theta/2)\mathbf{K}_{\zeta}$; similarly, the nearest-neighbors vectors on each layer read $\boldsymbol{\delta}_i^{(t,b)}=R(\pm \theta/2)\boldsymbol{\delta}_i$. Deviations of crystalline momentum, $\mathbf{p}=\mathbf{k}-\mathbf{K}_{\zeta}^{(t,b)}$, represent the conjugate variable to the position of the envelope wave functions, $\mathbf{p}\rightarrow -i\boldsymbol{\partial}$. Changes in the atomic arrangement within the moir\'e pitch are incorporated as a modulation of the crystal field parameter, $\Delta(\mathbf{r})$. We parametrize the stacking configurations by a vector field $\mathbf{d}(\rb)$ defined up to a microscopic lattice translation, see Fig.~\ref{fig:moire}. Thus, $\Delta(\mathbf{r})$ must be a functional of $\mathbf{d}(\mathbf{r})$ with the periodicity of the original Bravais lattice; a generic first-star expansion compatible with 3-fold symmetry reads
\begin{align}
    \label{eq:crystal_field}
    \Delta\left[\mathbf{d}\right]=\Delta_0+\sum_{i=1}^{3}\left[\Delta_1\cos\left(\mathbf{g}_i\cdot\mathbf{d}\right)+\Delta_2\sin\left(\mathbf{g}_i\cdot\mathbf{d}\right)\right].
\end{align}
In the moir\'e pattern, the stacking configuration varies as $\mathbf{d}(\mathbf{r})=2\sin\frac{\theta}{2}\,\mathbf{\hat{z}}\times\mathbf{r}+\mathbf{u}$. The modulation on the scale of the moir\'e pitch, $L_{\textrm{m}}=a/(2\sin\frac{\theta}{2})$, follows from the identity $\mathbf{g}_i\cdot\mathbf{d}(\mathbf{r})=\mathbf{G}_i\cdot(\mathbf{r}-\mathbf{\tilde{u}})$, where $\mathbf{\tilde{u}}=\mathbf{\hat{z}}\times\mathbf{u}/(2\sin\frac{\theta}{2})$ represents the center of the beating pattern corresponding to a AA' or AA local stacking. The parameters $\Delta_{\{0,1,2\}}$ can be extracted from fittings to band structure calculations of hBN bilayers in high-symmetry stackings \cite{portu_hBN}; we take $\Delta_0=4.10$~eV, $\Delta_{1}=-0.01$~eV, and $\Delta_2=0.08$~eV for type-I structures, and $\Delta_0=4.15$~eV, $\Delta_{1}=-0.02$~eV, and $\Delta_2=0$ (due to symmetry, see Fig.~\ref{fig:moire}) for type II.

For simplicity, we retain only interlayer scattering events between atoms sitting on top of each other conserving crystalline momentum. For each type of moir\'e structure, we consider three different parameters characterizing interlayer hoppings in boron-boron, nitrogen-nitrogen, and boron-nitrogen dimers. The interlayer tunneling matrix acquires the following structure:
\begin{subequations}
\label{eq:tunneling}
\begin{align}
   & \hat{T}_{\zeta}\left(\mathbf{r}\right)=\sum_{n=0}^{2}\hat{T}_{\zeta}^{(n)}e^{i\zeta\mathbf{k}_n\cdot\left(\mathbf{r}-\mathbf{\tilde{u}}\right)},\,\,\text{with}\\
   & \hat{T}_{\zeta}^{(n)}=e^{i\zeta\frac{n2\pi}{3}\hat{\sigma}_z}\, \hat{T}_0\,e^{-i\zeta\frac{n2\pi}{3}\hat{\sigma}_z},
\end{align} 
\end{subequations}
and $\mathbf{k}_n=R(n2\pi/3)\mathbf{k}_0$, with $\mathbf{k}_0=\mathbf{K}_+^{b}-\mathbf{K}_+^{t}$. For type-I structures, we obtain $\hat{T}_{0}=w_{\textrm{AA'}}\hat{1}+w_{\textrm{AB'}}\hat{\sigma}_++w_{\textrm{BA'}}\hat{\sigma}_-$, with $\hat{\sigma}_{\pm}=(\hat{\sigma}_x\pm i\hat{\sigma}_y)/2$ and $w_{\textrm{AA'}}=100$~meV, $w_{\textrm{AB'}}=300$~meV, and $w_{\textrm{BA'}}=80$~meV extracted from band structure calculations \cite{portu_hBN}. In the case of type-II structures, we have $\hat{T}_{0}=w_{\textrm{BB}}(\hat{1}+\hat{\sigma}_z)/2+w_{\textrm{NN}}(\hat{1}-\hat{\sigma}_z)/2+w_{\textrm{AB}}\,\hat{\sigma}_x$, where we take $w_{\textrm{BB}}=w_{\textrm{AB'}}$, $w_{\textrm{NN}}=w_{\textrm{BA'}}$, and $w_{\textrm{AB}}=200$~meV. The model is compatible with the $D_3$ symmetry of such structures, as shown in Fig.~\ref{fig:moire}~(c).

We diagonalize the Hamiltonian in a basis of (quasi-)Bloch states of the form 
\begin{align}
\label{eq:wf}
\psi_{\zeta,\mathbf{q}}^{\alpha,\mu}\left(\mathbf{r}\right)=\frac{e^{i\left(\mathbf{q}-\zeta\boldsymbol{\kappa}_{\mu}\right)\cdot\left(\mathbf{r}-\mathbf{\tilde{u}}\right)}}{\sqrt{A}}\sum_{\left\{\mathbf{G}\right\}}u_{\zeta,\mathbf{G}}^{\alpha,\mu}\left(\mathbf{q}\right)e^{i\mathbf{G}\cdot\left(\mathbf{r}-\mathbf{\tilde{u}}\right)},
\end{align}
where $A$ is the total area of the system and $\alpha$ ($\mu$) are the internal sublattice (layer) degrees of freedom. To avoid redundancies, quasi-momenta $\mathbf{q}$ must be restricted to the Brillouin zone of the moir\'e superlattice defined by the primitive beating pattern wavevectors, so that superpositions with $\mathbf{q}$ and $\mathbf{q}+\mathbf{G}$ represent the same electronic state; we impose periodic boundary conditions, $u_{\zeta,\mathbf{L}}^{\alpha,\mu}(\mathbf{q}+\mathbf{G})=u_{\zeta,\mathbf{L}+\mathbf{G}}^{\alpha,\mu}\left(\mathbf{q}\right)$. In this scheme, microscopic valleys are folded into the moir\'e Brillouin zone corners, $\mathbf{K}_{\pm}^{t}\equiv\pm\boldsymbol{\kappa}_t=\pm(2\mathbf{G}_2+\mathbf{G}_1)/3$ and $\mathbf{K}_{\pm}^{b}\equiv\pm\boldsymbol{\kappa}_b=\pm(2\mathbf{G}_1+\mathbf{G}_2)/3$. The sum in Eq.~\eqref{eq:wf} is truncated to a finite number of harmonics, $n_{\mathbf{G}}$. The coefficients $u_{\zeta,\mathbf{G}}^{\alpha,\mu}(\mathbf{q})$ can be arranged in a column vector, $|u_{\zeta,\mathbf{q}}\rangle$, defining a $4\,n_{\mathbf{G}}\times 4\,n_{\mathbf{G}}$ matrix to diagonalize for each $\mathbf{q}$. 

Narrow bands are formed in a broad range of twist angles, see Fig.~\ref{fig:bands}. We have chosen commensurate angles such that the moir\'e translation symmetry is exact and our results can be directly compared to microscopic calculations \cite{micro1,micro2}. The bandwidths of the lowest conduction and valence bands decrease with twist angle, ranging from $\sim200$~meV for $\theta=7.34^{\textrm{o}}$ to $\sim20$~meV for $\theta=3.89^{\textrm{o}}$; for $\theta\sim 3^{\textrm{o}}$, the bandwidths are only of a few meVs. We have included the full tight-binding dispersion of monolayer hBN. Linearizing $\phi(\mathbf{k})$ and neglecting the rotation of the sublattice basis does not change appreciably the bands. In that situation, the spectrum of twisted bilayer graphene recovers the electron-hole symmetry of the Dirac Hamiltonian \cite{Koshino}. For hBN, however, this symmetry is absent.

Panels~(c)~and~(d) show the particle density associated with the envelope wave functions of the lowest-energy flat bands at the $\boldsymbol{\gamma}$ point within a region containing four supercells [see Fig.~\ref{fig:moire}~(a1) for a comparison]. Profiles of similar symmetry are obtained at $\boldsymbol{\kappa}_{t,b}$ points. These electronic states are confined in regions of homopolar stacking, more localized as the bands enter deeper into the nominal gap of the commensurate structure. In type-I structures, the first conduction band is localized on boron sites of AB'-stacked regions, while the wave function of the first valence band is mostly on nitrogen sites of BA' stackings. In type-II structures, however, the wave functions are both localized in regions of AA stacking (in boron/nitrogen sites for conduction/valence band states). In both cases the wave function has equal weight on the top and bottom layer, due to C$_2$ symmetry. 

From these results, we posit that the emergence of flat bands in twisted bilayers of hBN and graphene responds to two different mechanisms of electron confinement. In hBN, the lowest flat bands result from resonant tunneling in stacking regions where atoms of the same polarity lie on top of each other. For graphene, however, a more subtle interference process takes place at certain (magic) twist angles. The most dramatic manifestation of the different nature of confinement takes place if we set the couplings $w_{\textrm{AB'}}=w_{\textrm{BA'}}= 0$: the lowest-energy flat bands of hBN disappear, whereas the analogous simplification in twisted bilayer graphene gives rise to completely flat bands at the magic angle \cite{chiral}.

\begin{figure}
\centerline{\includegraphics[width=\linewidth]{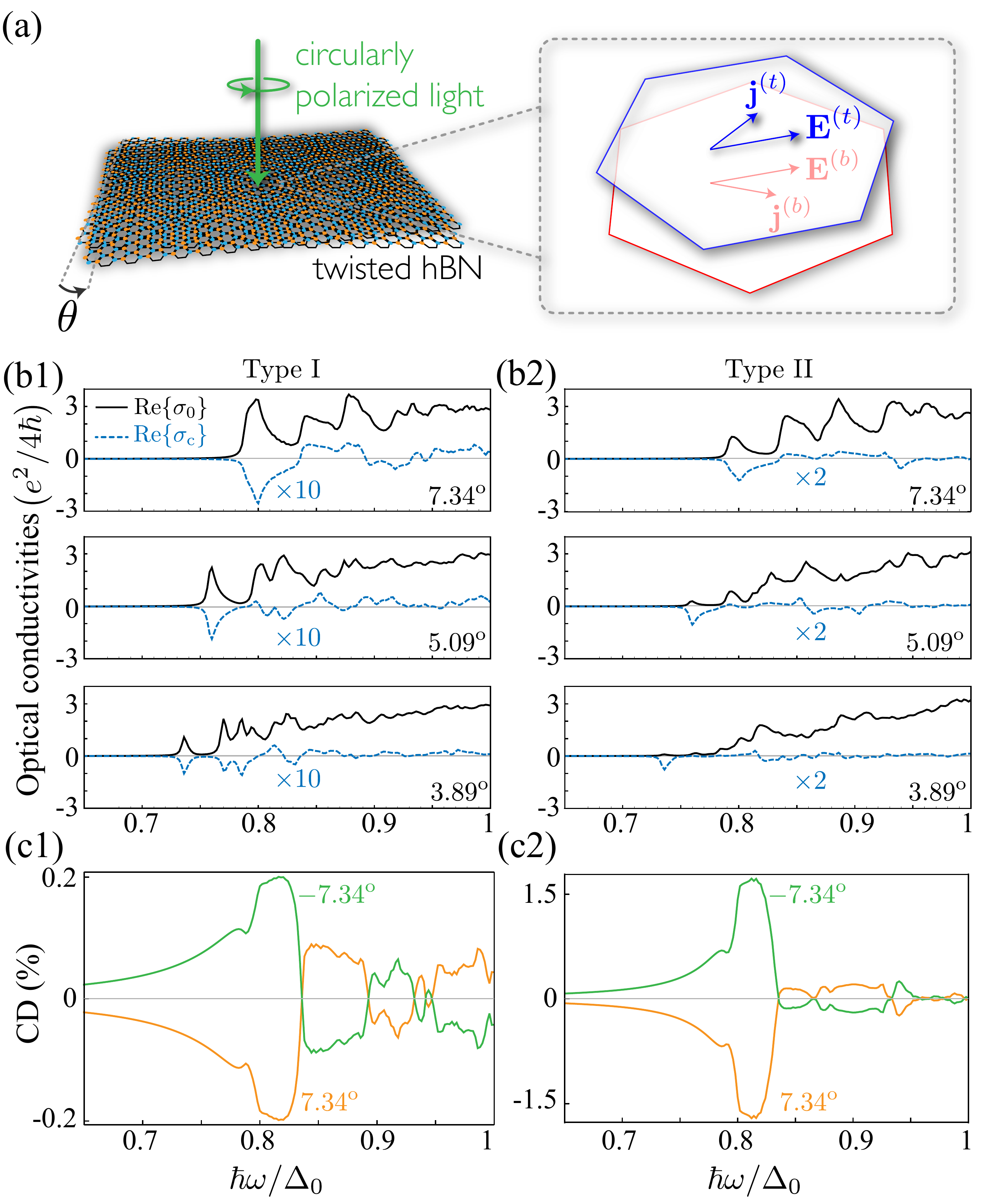}}
\caption{\textbf{Chiral optical response of twisted hBN.} (a) Incident light gives rise to (bound) currents that are not parallel to the electric field (subtending an angle proportional to $\sigma_\text{c}$). This generates a component of the magnetization collinear to the electric polarization, whose backaction rotates the polarization of light as it propagates through the material. The dissipative counterpart of this phenomenon  (known as ellipticity~\cite{Barron}) results in CD. (b) Real part of the total (black) and chiral (blue) conductivity in type I (b1) and type II (b2) moir\'e structures. The conductivities are expressed in units of the quantum of optical conductivity of graphene, $e^2/4\hbar$~\cite{ando}. The level broadening is $2\times 10^{-3}\Delta_0$. (c) Circular dichroism (in percentage) for structures with $|\theta|=7.34^{\textrm{o}}$ of type I (c1) and type II (c2).}
\label{fig:CD}
\end{figure}

The chirality (in real space) of the moir\'e structure has observable consequences in the optical response, leading to differential absorption of left- and right-circularly polarized light. Light (of frequency $\omega$ and at normal incidence) induces in-plane electrical currents, which can be decomposed in layer-resolved components $\mathbf{j}^{(t,b)}(\omega)$, as shown in Fig.~\ref{fig:CD}(a). The conjugate forces are the electric fields at the top and bottom layers, $\mathbf{E}^{(t,b)}(\omega)$. Following Ref.~\cite{Stauber}, we express currents and fields in terms of \textit{total} and \textit{counterflow} components, specifically, $\mathbf{j}=\mathbf{j}^{(t)}+\mathbf{j}^{(b)}$ and $\mathbf{j}^{(\textrm{cf})}=(\mathbf{j}^{(t)}-\mathbf{j}^{(b)})/2$ for the currents, and $\mathbf{E}=(\mathbf{E}^{(t)}+\mathbf{E}^{(b)})/2$ and $\mathbf{E}^{(\textrm{cf})}=\mathbf{E}^{(t)}-\mathbf{E}^{(b)}$ for the fields. Within this framework, the most general constitutive relations allowed by $D_3$ and time-reversal symmetries are\begin{align}
\label{eq:constitutive_relations}
  \left(\begin{array}{c}
    \mathbf{j}\\
    \mathbf{j}^{(\textrm{cf})}
    \end{array}\right)= \left[ \begin{array}{cc}
   \sigma_0\left(\omega\right) & \sigma_{\textrm{c}}\left(\omega\right)\mathbf{\hat{z}}\times\\
  - \sigma_{\textrm{c}}\left(\omega\right)\mathbf{\hat{z}}\times &  \sigma_{\textrm{cf}}\left(\omega\right)
    \end{array}\right]\left(\begin{array}{c}
    \mathbf{E}\\
    \mathbf{E}^{(\textrm{cf})}
    \end{array}\right).
\end{align}
Here $\sigma_{0(\textrm{cf})}\left(\omega\right)$ is the total (counterflow) conductivity, characterizing the longitudinal response of the system; $\sigma_{\textrm{c}}$ is the \textit{chiral} conductivity, accounting for the transverse flow of charge in one layer generated by the electric field in the other layer. The latter response arises from the chirality of the structure and governs the optical activity. 

We compute the total and chiral conductivities from the Kubo formula. The contributions from interband transitions are given by the regular part of current-current correlation functions with suitable combinations of coordinate and layer indices in each case \cite{SM}. The final result (including spin degeneracy) reads \begin{subequations}
\label{eq:conductivity}
\begin{align}
    & \sigma_{i}\left(\omega\right)=\sum_{n, m}\int\frac{d\mathbf{q}}{\left(2\pi\right)^2}\frac{4ie^2\omega\,\mathcal{O}_{i,+,\mathbf{q}}^{n,m}\left(f_{+,\mathbf{q}}^n-f_{+,\mathbf{q}}^m\right)}{\hbar\omega+i0^{+}+\varepsilon_{+,\mathbf{q}}^{n}-\varepsilon_{+,\mathbf{q}}^{m}},\\
    & \text{with}\,\,\,\, \mathcal{O}_{0,\zeta,\mathbf{q}}^{n,m}=\frac{\hbar^2|\braket{ u_{\zeta,\mathbf{q}}^n|\hat{v}_{\zeta}^{x}|u_{\zeta,\mathbf{q}}^m}|^2}{(\varepsilon_{\zeta,\mathbf{q}}^{n}-\varepsilon_{\zeta,\mathbf{q}}^{m})^2},\,\,\,\text{and}\\
    & \mathcal{O}_{\textrm{c},\zeta,\mathbf{q}}^{n,m}=\frac{\hbar^2\textrm{Re}\left\{\braket{u_{\zeta,\mathbf{q}}^n|\hat{v}_{\zeta,t}^{x}|u_{\zeta,\mathbf{q}}^m}\braket{u_{\zeta,\mathbf{q}}^m|\hat{v}_{\zeta,b}^{y}|u_{\zeta,\mathbf{q}}^n}\right\}}{(\varepsilon_{\zeta,\mathbf{q}}^{n}-\varepsilon_{\zeta,\mathbf{q}}^{m})^2}.
\end{align}
\end{subequations}
In these expression $f_{+,\mathbf{q}}^n$ are the Fermi-Dirac occupation numbers for (quasi-)Bloch electrons in band $n$ with quasi-momentum $\mathbf{q}$ within valley $\zeta=+1$ and $\varepsilon_{+,\mathbf{q}}^n$ is the corresponding band energy. We have also introduced \textit{oscillator strengths} containing matrix elements of the total velocity, $\boldsymbol{\hat{v}}_{\zeta}\equiv(-i/\hbar)[\hat{\mathbf{r}},\hat{\mathcal{H}}_{\zeta}]$, and layer-resolved velocity operators, $\boldsymbol{\hat{v}}_{\zeta,\mu}\equiv \hat{P}_{\mu}\boldsymbol{\hat{v}}_{\zeta}\hat{P}_{\mu}$, where $\hat{P}_{\mu}$ are projectors on layer subspace labeled by $\mu$. The sum is on band indices and the integration is over the moir\'e Brillouin zone.

Sharp resonances in the real part of the total conductivity appear within the nominal gap of hBN, as shown in Fig.~\ref{fig:CD}~(b). They are associated with optical transitions between the low-energy flat bands. The chiral conductivity in type-II structures is comparable (and even larger for small twist angles) to the total conductivity. This is manifested in the CD, which is defined as
\cite{Barron}\begin{align}
\label{eq:ellipticity}
    \text{CD}=\frac{\mathcal{A}_+-\mathcal{A}_-}{(\mathcal{A}_++\mathcal{A}_-)/2}\approx \frac{\omega z_0}{c}\frac{4\,\textrm{Re}\,\sigma_{\textrm{c}}}{\textrm{Re}\,\sigma_{0}},
\end{align}
where $\mathcal{A}_{+(-)}$ is the absorbance of right- (left-) circularly polarized light~\cite{SM}. In the above expression $z_0\approx 3.6$~ \AA\, is the interlayer distance and $c$ is the speed of light. Figure~\ref{fig:CD}~(c) shows the CD of chiral partners with twist angles $|\theta|=7.34^{\textrm{o}}$. In both types of structures, the CD is resonant with the transition between the lowest-energy flat bands and is dominated by electronic states around the $\boldsymbol{\gamma}-\boldsymbol{m}$ axes, where interlayer hibridizations are maximized. The CD is enhanced by decreasing the twist angle, but at the expense of the overall absorption signal.

Surprisingly, the two types of moir\'es give rise to  significantly different chiral responses. The circular dichroism of type-I structures is of the same order as in twisted bilayer graphene ($\sim20$~mdeg at the first resonance) \cite{CD_exp,Brey}, but in type II is one order of magnitude larger. The reason is that interlayer current correlations defining the chiral conductivity are mostly determined by the hopping between atoms of opposite polarity, which is larger in type-II structures. Setting $w_{\textrm{AA'}}=0$ or $w_{\textrm{AB}}=0$ in each case does not alter significantly the electronic spectrum but suppresses the CD. This can be traced back to the emergence of approximate \textit{pseudo-inversion} symmetries in these limits, which leads to cancellations in $\sigma_{\textrm{c}}$ coming from interband processes at opposite points in reciprocal space \cite{SM}. We stress that these are not spatial symmetries (i.e., not inherited from the structure), but instead arise from the coherences between the sublattice and layer degrees of freedom in the electronic wave function that govern the chiral response. A similar observation has been recently made for twisted bilayer graphene at the magic angle, where the cancellation of the (intraband) chiral response comes together with an emergent sixfold symmetry~\cite{stauber-2}. All these results suggest that, in general, there is no one-to-one correspondence between the chirality of the structure (i.e., the twist direction) and the chirality of the optical response.

In conclusion, twisted bilayers of polar materials host flat bands without the requirement of a special magic-angle alignment. These bands are built from in-gap resonant states at regions of unnatural stacking, where atoms of the same species lie on top of each other. In the case of hBN, the reconstruction of the electronic spectrum is manifested as sharp resonances in the optical absorption spectra deep inside the original gap. Owing to their chiral structure, these materials absorb left and right circularly-polarized light differently. This CD originates from current/polarization correlations between the layers, generated mostly in regions of heteropolar stacking. Our predictions for hBN are within experimental reach, as the formation of flat bands contribute to a 25\%-reduction of the optical gap. The CD in the ultraviolet can be used not only to identify the twist angle, but also to distinguish between the two types of moir\'e structures with different symmetry. At lower frequencies, the response will be dominated by phonons, excitons and other hybrid quasiparticles. Their contribution to the chiral optical response is also an interesting subject for future research. Finally, the sensitivity to stacking reported here also lays out the question of how commensuration effects at large angles \cite{mele} and lattice relaxation at small angles affect this observable.

\vspace{10pt}
\textbf{Acknowledgments.}-- The authors thank D. Basov, A. N. Pasupathy, and P. J. Schuck for useful discussions. This work is financially supported by Programmable Quantum Materials, an Energy Frontier Research Center funded by the U.S. Department of Energy (DOE), Office of Science, Basic Energy Sciences (BES), under award DE-SC0019443.

\clearpage
\onecolumngrid

\appendix

\section{Supplemental Information: Flat bands and chiral optical response of moir\'e insulators}
We derive Eqs.~(6) of the main text from the Kubo formula and the leading contribution to the circular dichroism in Eq.~(7). We analyze the presence of approximate \textit{pseudo-inversion} symmetries in certain limits of the model, leading to cancellations in the circular dichroism. 
\maketitle
\onecolumngrid

\subsection{Calculation of the optical conductivity}

\subsubsection{Hamiltonian}

The Hamiltonian can be written as $\hat{H}=\hat{H}_0+\hat{V}(t)$, where $\hat{H}_0$ describes the stationary electronic spectrum and $\hat{V}(t)$ represents the coupling with the external field. The goal is to compute the response of the system to linear order in the perturbation $\hat{V}(t)$.

In the continuum model, the first term can be written as\begin{align}
    \hat{H}_0=\sum_{\zeta=\pm 1}\int d\mathbf{r}\, \hat{\psi}_{\zeta}^{\dagger}\left(\mathbf{r}\right)\hat{\mathcal{H}}_{\zeta}\left(\mathbf{r}\right) \hat{\psi}_{\zeta}\left(\mathbf{r}\right),
\end{align}
where $\hat{\psi}_{\zeta}\left(\mathbf{r}\right)$ is the electron field operator in valley $\zeta$. Note that $\hat{\psi}_{\zeta}^{\dagger}\left(\mathbf{r}\right)$ is in fact a 4-column vector containing fermion operators that create an electron in position $\mathbf{r}$ of sublattice $\alpha$ and layer $\mu$. In the continuum model, these operators are defined by a $\mathbf{k}\cdot\mathbf{p}$ expansion around valley $\zeta$ of the microscopic field operator given by\begin{align}
    \hat{\Psi}\left(\mathbf{r}\right)=\sum_{\alpha,\mu}\sum_{\zeta=\pm 1}f_{\zeta}^{\alpha,\mu}\left(\mathbf{r}\right)\,\hat{\psi}_{\zeta}^{\alpha,\mu}\left(\mathbf{r}\right),
\end{align}
where $f_{\zeta}^{\alpha,\mu}(\mathbf{r})$ represents the fast (on the atomic scale) oscillating factors of the electronic wave function. In our tight-biding description of hBN, these fast oscillating factors are nothing but the Bloch wave functions at $\mathbf{K}_{\zeta}^{(\mu)}$ points, \begin{align}
f_{\zeta}^{\alpha,\mu}(\mathbf{r})=\frac{1}{\sqrt{N}}\sum_{i}e^{i\mathbf{K}_{\zeta}^{(\mu)}\cdot\mathbf{R}_{i}^{\alpha,\mu}}\,\Phi^{\alpha,\mu}\left(\mathbf{r}-\mathbf{R}_{i}^{\alpha,\mu}\right),
\end{align}
where $\mathbf{R}_{i}^{\alpha,\mu}$ represent the positions of the atoms, the sum is extended to $N$ microscopic cells, and $\Phi^{\alpha,\mu}(\mathbf{r})$ are Wannier functions of $\pi$ orbitals in $\alpha$ sublattice of layer $\mu$.

The continuum Hamiltonian in the main text is written in a gauge of envelope wave functions such that the translational invariance is defined only up to a unitary rotation, namely,\begin{align}
\label{eq:translation_symmetry}
\hat{\mathcal{H}}_{\zeta}\left(\mathbf{r}+\mathbf{R}\right)= \hat{\mathcal{U}}_{\zeta}\left(\mathbf{R}\right)\, \hat{\mathcal{H}}_{\zeta}\left(\mathbf{r}\right) \, \hat{\mathcal{U}}_{\zeta}^{\dagger}\left(\mathbf{R}\right),\,\,\,\text{with}\,\,\,\hat{\mathcal{U}}\left(\mathbf{R}\right)=\left[\begin{array}{cc}
 e^{-i\zeta\boldsymbol{\kappa}_t\cdot\mathbf{R}}\,\hat{1} & 0\\
0 & e^{-i\zeta\boldsymbol{\kappa}_b\cdot\mathbf{R}}\,\hat{1}
\end{array}\right].
\end{align}
Here $\mathbf{R}$ is a vector of the moir\'e superlattice such that $e^{i\mathbf{G}_i\cdot\mathbf{R}}=1$. The unitary operator in layer space contains momentum boosts associated with the shift of the positions of the microscopic valleys. Bloch wave functions in this gauge satisfy\begin{align}
\label{eq:Bloch_theorem}
\psi_{\zeta,\mathbf{q}}^{\alpha,\mu}\left(\mathbf{r}+\mathbf{R}\right)=e^{i\left(\mathbf{q}-\zeta\boldsymbol{\kappa}_{\mu}\right)\cdot\mathbf{R}}\,\psi_{\zeta,\mathbf{q}}^{\alpha,\mu}\left(\mathbf{r}\right),
\end{align} 
with $\mathbf{q}$ restricted to the first Brillouin zone of the moir\'e superlattice. This form of Bloch theorem leads to the ansatz in Eq.~(4) of the main text. The extra phase factor may look odd at first glance, but note that is just a reminiscence of the fast oscillating factor of the microscopic wave function since $f_{\zeta}^{\alpha,\mu}(\mathbf{r}+\mathbf{R})\approx e^{i\mathbf{K}_{\zeta}^{(\mu)}\cdot\mathbf{R}}\, f_{\zeta}^{\alpha,\mu}(\mathbf{r})$, where the identity is exhausted only when the moir\'e superlattice symmetry is microscopically exact. In that case, the folding scheme imposing $\mathbf{K}_{\zeta}^{(\mu)}\equiv \zeta\boldsymbol{\kappa}_{\mu}$ is exact and the microscopic wave function $\Psi_{\mathbf{q}}^{\alpha,\mu}(\mathbf{r})=\sum_{\zeta}f_{\zeta}^{\alpha,\mu}(\mathbf{r})\times\psi_{\zeta,\mathbf{q}}^{\alpha,\mu}(\mathbf{r})$ satisfies the Bloch theorem in the usual form, $\Psi_{\mathbf{q}}^{\alpha,\mu}(\mathbf{r}+\mathbf{R})=e^{i\mathbf{q}\cdot\mathbf{R}}\,\Psi_{\mathbf{q}}^{\alpha,\mu}(\mathbf{r})$.

The translation of this discussion to second quantization language is that the band Hamiltonian is diagonalized in a basis of operators of the form\begin{align}
    \hat{c}_{\zeta}^{\alpha,\mu}\left(\mathbf{k}\right)=\frac{1}{\sqrt{A}}\int d\mathbf{r}\, e^{-i\left(\mathbf{k}-i\zeta\boldsymbol{\kappa}_{\mu}\right)\cdot\mathbf{r}}\,\hat{\psi}_{\zeta}^{\alpha,\mu}\left(\mathbf{r}\right).
\end{align}
In this definition, momentum $\mathbf{k}$ is not restricted to the first Brillouin zone of the moir\'e superlattice, but it is convenient to introduce such restriction in the expression of the Hamiltonian. In order to do so, let us introduce the following ket notation:\begin{subequations}\begin{align}
&  \left|\mathbf{q}\right\rangle\otimes \left|\zeta,\alpha,\mu,\mathbf{G}\right\rangle\equiv \left[\hat{c}_{\zeta}^{\alpha,\mu}\left(\mathbf{q}+\mathbf{G}\right)\right]^{\dagger}\left|0\right\rangle, \\
& \text{such that}\,\,\, \left\langle \mathbf{r}\right|\mathbf{q}\rangle=\frac{e^{i\mathbf{q}\cdot\mathbf{r}}}{\sqrt{N_{\textrm{m}}}},\,\,\text{and}\\
& \left\langle \mathbf{r}\right|\zeta,\alpha,\mu,\mathbf{G}\rangle=\frac{e^{i\left(\mathbf{G}-\zeta\boldsymbol{\kappa}_{\mu}\right)\cdot\mathbf{r}}}{\sqrt{A_{\textrm{m}}}},
\end{align}\end{subequations}
where $N_{\textrm{m}}$ is the number of moir\'e supercells and $A_{\textrm{m}}$ is their area. Here we have just written momentum $\mathbf{k}$ as $\mathbf{k}=\mathbf{q}+\mathbf{G}$, with $\mathbf{q}$ restricted to the first moir\'e Brillouin zone (defining the usual Bloch factor), and the Fourier replica in the reciprocal lattice $\mathbf{G}$ has been incorporated as a new \textit{internal} quantum number. The Hamiltonian reads then\begin{align}
\hat{H}_0=\sum_{\zeta=\pm 1}\sum_{\mathbf{q}\in\textrm{mBZ}}\hat{\mathcal{H}}_{\zeta}\left(\mathbf{q}\right)\otimes \left|\mathbf{q}\right\rangle\left \langle \mathbf{q}\right|,
\end{align}
where $\hat{\mathcal{H}}_{\zeta}\left(\mathbf{q}\right)$ is a $4n_{\mathbf{G}}\times 4n_{\mathbf{G}}$ matrix of the form (in the case of type-II structures)\begin{align}
\nonumber
\hat{\mathcal{H}}_{\zeta}\left(\mathbf{q}\right)=&\sum_{\alpha,\beta}\sum_{\left\{\mathbf{G}\right\}} \sum_{\mu} \left[t\,\textrm{Re}\left\{\phi_{\mu}\left(\mathbf{q}-\zeta\boldsymbol{\kappa}_{\mu}+\mathbf{G}\right)\right\}\hat{\sigma}_{x}+t\,\textrm{Im}\left\{\phi_{\mu}\left(\mathbf{q}-\zeta\boldsymbol{\kappa}_{\mu}+\mathbf{G}\right)\right\}\hat{\sigma}_{y}+\frac{\Delta_0}{2}\hat{\sigma}_z\right]_{\alpha\beta} \left|\zeta,\alpha,\mu,\mathbf{G}\right\rangle\left\langle \zeta,\beta,\mu,\mathbf{G} \right|\\
& +\frac{\Delta_1-i\Delta_2}{4}\sum_{\alpha,\beta}\sum_{\left\{\mathbf{G}\right\}}\sum_{\mu}\sum_{i=1,2,3} \left[\hat{\sigma}_z\right]_{\alpha\beta}\left|\zeta,\alpha,\mu,\mathbf{G}+\mathbf{G}_i\right\rangle\left\langle \zeta,\beta,\mu,\mathbf{G} \right| + \textrm{h.c.}\\
& +\sum_{\alpha,\beta}\sum_{\left\{\mathbf{G}\right\}}\sum_{n=0,1,2} \left[\hat{T}_{\zeta}^{(n)}\right]_{\alpha\beta}\left|\zeta,\alpha,t,\mathbf{G}+\zeta\mathbf{Q}_n\right\rangle\left\langle \zeta,\beta,b,\mathbf{G} \right| + \textrm{h.c.},
\nonumber
\end{align}
with $\mathbf{Q}_0=\mathbf{0}$, $\mathbf{Q}_1=\mathbf{G}_2$, $\mathbf{Q}_2=-\mathbf{G}_1$ and\begin{align}
\label{eq:phi}
    \phi_{\mu}\left(\mathbf{p}\right)\equiv\sum_{i=1,2,3}e^{i\boldsymbol{\delta}_i^{(\mu)}\cdot\left(\mathbf{p}+\mathbf{K}_{\zeta}^{(\mu)}\right)}.
\end{align}
In the case of type-I structures, note that the paramters $\Delta_i$ enter with opposite sign in the bottom layer sector. The solution of the diagonalization problem of this matrix defines the band energies $\varepsilon_{\zeta,\mathbf{q}}^n$ and the coefficients $u_{n,\zeta,\mathbf{G}}^{\alpha,\mu}\left(\mathbf{q}\right)$ in the ansatz of Eq.~(4) of the main text, which are just the $4n_{\mathbf{G}}$ components of the corresponding eigenvector:\begin{align}
\left|u_{n,\zeta}\left(\mathbf{q}\right) \right\rangle= \sum_{\alpha}\sum_{\mu}\sum_{\left\{ \mathbf{G}\right\}} u_{n,\zeta,\mathbf{G}}^{\alpha,\mu}\left(\mathbf{q}\right)\left|\zeta,\alpha,\mu,\mathbf{G}\right\rangle.
\end{align}
In second quantization, this last relation amounts to \begin{align}
\label{eq:relation_operators}
    \hat{c}_{\zeta}^{\alpha,\mu}\left(\mathbf{q}+\mathbf{G}\right)=\sum_n u_{n,\zeta,\mathbf{G}}^{\alpha,\mu}\left(\mathbf{q}\right)\, \hat{c}_{n,\zeta}\left(\mathbf{q}\right),
\end{align}
where the summation is in band indices. Here $\hat{c}_{n,\zeta}\left(\mathbf{q}\right)$ are fermion operators associated with excitations in band $n$ of valley $\zeta$ with quasi-momentum $\mathbf{q}$,\begin{align}
    \hat{c}_{\zeta,n}^{\dagger}\left(\mathbf{q}\right)\left|0\right\rangle=\left|\mathbf{q}\right\rangle\otimes \left|u_{n,\zeta}\left(\mathbf{q}\right) \right\rangle.
\end{align}

The relation between operators in Eq.~\eqref{eq:relation_operators} is univocal as long as we define boundary conditions in reciprocal space. Note that if an infinite number of Fourier replica $n_{\mathbf{G}}$ is included, then the matrix Hamiltonian is periodic in the momentum space up to a unitary transformation,\begin{align}
\hat{\mathcal{H}}_{\zeta}\left(\mathbf{q}+\mathbf{G}_i\right)=\hat{\mathcal{U}}_{\mathbf{G}_i}\hat{\mathcal{H}}_{\zeta}\left(\mathbf{q}\right)\hat{\mathcal{U}}_{\mathbf{G}_i}^{\dagger}.
\end{align}
This unitary transformation reads\begin{align}
\hat{\mathcal{U}}_{\mathbf{G}_i}=\sum_{\alpha}\sum_{\mu}\sum_{\left\{ \mathbf{G}\right\}} \left|\zeta,\alpha,\mu,\mathbf{G}-\mathbf{G}_i\right\rangle\left\langle \zeta,\alpha,\mu,\mathbf{G} \right|.
\end{align}
It follows that if $|u_{n,\zeta}(\mathbf{q}) \rangle$ is an eigenvector of $\hat{\mathcal{H}}_{\zeta}(\mathbf{q})$ with eigenvalue $\varepsilon_{\zeta,\mathbf{q}}^n$, then $\hat{\mathcal{U}}_{\mathbf{G}_i}|u_{n,\zeta}(\mathbf{q}) \rangle$ is an eigenvector of $\hat{\mathcal{H}}_{\zeta}(\mathbf{q}+\mathbf{G}_i)$ with the same eigenvalue. Periodic boundary conditions amounts to the identification\begin{align}
\left|u_{n,\zeta}\left(\mathbf{q}+\mathbf{G}_i\right) \right\rangle\equiv \hat{\mathcal{U}}_{\mathbf{G}_i} \left|u_{n,\zeta}\left(\mathbf{q}\right) \right\rangle \Longrightarrow u_{n,\zeta,\mathbf{G}}^{\alpha,\mu}\left(\mathbf{q}+\mathbf{G}_i\right)=u_{n,\zeta,\mathbf{G}+\mathbf{G}_i}^{\alpha,\mu}\left(\mathbf{q}\right),\,\, \hat{c}_{n,\zeta}\left(\mathbf{q}+\mathbf{G}_i\right)=\hat{c}_{n,\zeta}\left(\mathbf{q}\right).
\end{align}

Finally, as the continuum model is derived from a tight-binding Hamiltonian, it is natural to introduce the coupling with external fields via a Peierls substitution, $\hbar\mathbf{q}\rightarrow \hbar\mathbf{q}-e\mathbf{A}\left(t,\mathbf{r},z\right)$. Before writing down the perturbation, some remarks are in order. First, in our model, the $z$ coordinate only enters through the internal \textit{layer} number, so instead of dealing with a function of $z$, we should consider the values of the external field in the top and bottom layer,\begin{align}
    \mathbf{A}^{(t,b)}\left(t,\mathbf{r}\right)\equiv \mathbf{A}\left(t,\mathbf{r},\pm\frac{z_0}{2}\right).
\end{align}
Second, as the model only retains interlayer hopping processes between atoms sitting on top of each other and these processes do not carry lateral currents, the total miscroscopic currents can be also decomposed into two layer-resolved components. To the leading order in the external field, the perturbation reads\begin{align}
    \hat{V}\left(t\right)=-\sum_{\mu}\int d\mathbf{r}\,\,\mathbf{\hat{j}}_{p}^{(\mu)}\left(\mathbf{r}\right)\cdot\mathbf{A}^{(\mu)}\left(t,\mathbf{r}\right)=-\sum_{\mu}\sum_{\mathbf{k}}\,\mathbf{\hat{j}}_{p}^{(\mu)}\left(\mathbf{k}\right)\cdot\mathbf{A}_{\mathbf{k}}^{(\mu)}\left(t\right),
\end{align}
where $\mathbf{A}_{\mathbf{k}}^{(\mu)}\left(t\right)$ are the Fourier components of the external field,\begin{align}
    \mathbf{A}_{\mathbf{k}}^{(\mu)}\left(t\right)=\frac{1}{\sqrt{A}}\int d\mathbf{r}\,\mathbf{A}^{(\mu)}\left(t,\mathbf{r}\right) e^{-i\mathbf{k}\cdot\mathbf{r}},
\end{align}
and $\mathbf{\hat{j}}_{p}^{(\mu)}\left(\mathbf{k}\right)$ is the paramagnetic term of the layer-resolved current operator in momentum representation,\begin{align}
\mathbf{\hat{j}}_{p}^{(\mu)}\left(\mathbf{k}\right)=\frac{e}{\sqrt{A}}\sum_{n1,n_2}\sum_{\zeta}\sum_{\mathbf{q}\in\textrm{mBZ}}\left\langle u_{n_1,\zeta}\left(\mathbf{k}+\mathbf{q}\right)\right| \boldsymbol{\hat{v}}_{\zeta,\mu} \left|u_{n_2,\zeta}\left(\mathbf{q}\right) \right\rangle\, \hat{c}_{n_1,\zeta}^{\dagger}\left(\mathbf{k}+\mathbf{q}\right)\hat{c}_{n_2,\zeta}\left(\mathbf{q}\right).
\end{align}
In this last expression we have used the relation in Eq.~\eqref{eq:relation_operators} along with the representation in momentum space of the (layer-resolved) velocity operator, which is just\begin{align}
    \boldsymbol{\hat{v}}_{\zeta}=\frac{1}{\hbar}\partial_{\mathbf{q}}\hat{\mathcal{H}}_{\zeta}\left(\mathbf{q}\right).
\end{align}
Next, we elaborate a bit more on the structure of the current operators before addressing the calculation in the last subsection.

\subsubsection{Current operators in the continuum model}

The layer-resolved density operators can be defined as usual in terms of field operators of the continuum model, \begin{align}
    \hat{\rho}^{(\mu)}\left(\mathbf{r}\right)=e\sum_{\zeta}\sum_{\alpha}\left[\hat{\psi}_{\zeta}^{\alpha,\mu}\left(\mathbf{r}\right)\right]^{\dagger}\hat{\psi}_{\zeta}^{\alpha,\mu}\left(\mathbf{r}\right)=\frac{1}{\sqrt{A}}\sum_{\mathbf{k}}e^{-i\mathbf{k}\cdot\mathbf{r}}\hat{\rho}^{(\mu)}\left(\mathbf{k}\right),
\end{align}
where \begin{align}
    \hat{\rho}^{(\mu)}\left(\mathbf{k}\right)=\frac{e}{\sqrt{A}}\sum_{\zeta}\sum_{\alpha}\sum_{\mathbf{p}}\left[\hat{c}_{\zeta}^{\alpha,\mu}\left(\mathbf{k}+\mathbf{p}\right)\right]^{\dagger}\hat{c}_{\zeta}^{\alpha,\mu}\left(\mathbf{p}\right).
\end{align}
Note that the summations in momentum are not restricted in either of the last two expressions. Introducing the relations in Eq.~\eqref{eq:relation_operators} and the ket notation discussed before, we arrive at the compact expression\begin{align}
    \hat{\rho}^{(\mu)}\left(\mathbf{k}\right)=\frac{e}{\sqrt{A}}\sum_{n_1,n_2}\sum_{\zeta}\sum_{\mathbf{q}\in\textrm{mBZ}}\left\langle u_{n_1,\zeta}\left(\mathbf{k}+\mathbf{q}\right)\right| \hat{P}_{\mu} \left|u_{n_2,\zeta}\left(\mathbf{q}\right) \right\rangle\, \hat{c}_{n_1,\zeta}^{\dagger}\left(\mathbf{k}+\mathbf{q}\right)\hat{c}_{n_2,\zeta}\left(\mathbf{q}\right).
\end{align}

Charge continuity in momentum space implies\begin{align}
\sum_{\mu}\dot{\hat{\rho}}^{(\mu)}\left(\mathbf{k}\right)=\frac{i}{\hbar}\left[\hat{H},\sum_{\mu}\hat{\rho}^{(\mu)}\left(\mathbf{k}\right)\right]=i\sum_{\mu}\mathbf{k}\cdot\mathbf{\hat{j}}^{(\mu)}\left(\mathbf{k}\right),
\end{align}
where $\mathbf{\hat{j}}^{(\mu)}$ the total (paramagnetic plus diamagnetic) layer-resolved current. The commutator reads (expanding the band Hamiltonian up to first order)\begin{align}
\frac{i}{\hbar}\left[\hat{H},\sum_{\mu}\hat{\rho}^{(\mu)}\left(\mathbf{k}\right)\right]=i\sum_{\mu}\mathbf{k}\cdot\mathbf{\hat{j}}_p^{(\mu)}\left(\mathbf{k}\right)-\frac{i}{\hbar}\sum_{\mu,\nu}\sum_{\mathbf{p}}\mathbf{A}_{\mathbf{p}}^{(\nu)}\left(t\right)\cdot\left[\mathbf{\hat{j}}_p^{(\nu)}\left(\mathbf{p}\right),\hat{\rho}^{(\mu)}\left(\mathbf{k}\right)\right].
\end{align}
The commutator in the second term of the right-hand side of the last equation reads (up to the same order in external momentum)\begin{align}
\left[\mathbf{\hat{j}}_p^{(\nu)}\left(\mathbf{p}\right),\hat{\rho}^{(\mu)}\left(\mathbf{k}\right)\right]=\frac{e^2\hbar \,k_i\,\mathbf{\hat{e}}_j}{A}\sum_{n_1,n_2}\sum_{\zeta}\sum_{\mathbf{q}\in\textrm{mBZ}}\left\langle u_{n_1,\zeta}\left(\mathbf{p}+\mathbf{k}+\mathbf{q}\right)\right| \hat{P}_{\nu} \hat{\mathcal{D}}_{\zeta}^{i,j} \hat{P}_{\mu} \left|u_{n_2,\zeta}\left(\mathbf{q}\right) \right\rangle\,
\hat{c}_{n_1,\zeta}^{\dagger}\left(\mathbf{p}+\mathbf{k}+\mathbf{q}\right) \hat{c}_{n_2,\zeta}\left(\mathbf{q}\right),
\end{align}
where summation over spatial (latin) indices is assumed hereafter and we have introduce the inverse effective-mass tensor,\begin{align}
    \hat{\mathcal{D}}_{\zeta}^{i,j}=\frac{1}{\hbar^2}\partial_{q_i}\partial_{q_j}\hat{\mathcal{H}}_{\zeta}\left(\mathbf{q}\right).
\end{align}
From this last result we can identify the total layer-resolved current operator $\mathbf{\hat{j}}^{\mu}=\mathbf{\hat{j}}_p^{\mu}+\mathbf{\hat{j}}_d^{\mu}$, where the diamagnetic term reads\begin{align}
\mathbf{\hat{j}}_d^{\mu}\left(\mathbf{k}\right)=-\frac{e^2\mathbf{\hat{e}}_i}{A}\sum_{n_1,n_2}\sum_{\zeta}\sum_{\mathbf{p}}\sum_{\mathbf{q}\in\textrm{mBZ}}\left(\mathbf{\hat{e}}_{j}\cdot \mathbf{A}_{\mathbf{p}}^{(\mu)}\left(t\right)\right)\left\langle u_{n_1,\zeta}\left(\mathbf{k}+\mathbf{p}+\mathbf{q}\right)\right| \hat{\mathcal{D}}_{\zeta,\mu}^{i,j} \left|u_{n_2,\zeta}\left(\mathbf{q}\right) \right\rangle\hat{c}_{n_1,\zeta}^{\dagger}\left(\mathbf{k}+\mathbf{p}+\mathbf{q}\right) \hat{c}_{n_2,\zeta}\left(\mathbf{q}\right),
\end{align}
where we have used $\hat{P}_{\mu}\hat{\mathcal{D}}_{\zeta}^{i,j}\hat{P}_{\nu}=\hat{P}_{\mu}\hat{\mathcal{D}}_{\zeta}^{i,j}\hat{P}_{\mu}\delta_{\mu,\nu}$ since the operator is diagonal in layer indices and $\hat{\mathcal{D}}_{\zeta,\mu}^{i,j} \equiv \hat{P}_{\mu}\hat{\mathcal{D}}_{\zeta}^{i,j}\hat{P}_{\mu}$. Note, however, that the diamagnetic term only enters in the intraband contribution to the total conductivity. If we retained linear terms in $\mathbf{q}$ only, the diamagnetic term would be exactly zero.

\subsubsection{Kubo formulas}

We proceed now to the perturbative calculation. Let us consider normal incidence and assume that the light spot is large compared to the moir\'e pitch. We work in the Weyl gauge and assume the dipolar approximation for the incident field, $\mathbf{A}(t,\mathbf{r})^{(\mu)}=\mathbf{A}^{(\mu)}(t)$, such that $\mathbf{E}^{(\mu)}(t)=-\dot{\mathbf{A}}^{(\mu)}(t)$. As customary, we will replace the incident fields by the total fields at the end of the calculation, neglecting local field corrections generated by induced currents.

Under these assumptions, the perturbation contains only the $\mathbf{k}=0$ component of the paramagnetic current. In frequency domain, we just have\begin{align}
  \hat{V}\left(\omega\right)=\frac{ie}{\omega}\sum_{\mu}\sum_{n_1,n_2}\sum_{\zeta}\sum_{\mathbf{q}\in\textrm{mBZ}}\left\langle u_{n_1,\zeta}\left(\mathbf{q}\right)\right|\mathbf{E}^{\mu}\left(\omega\right)\cdot\boldsymbol{\hat{v}}_{\zeta,\mu}  \left | u_{n_2,\zeta}\left(\mathbf{q}\right) \right\rangle\, \hat{c}_{n_1,\zeta}^{\dagger}\left(\mathbf{q}\right) \hat{c}_{n_2,\zeta}\left(\mathbf{q}\right).
\end{align}
We want to compute the linear response in the total current, provided that the latter is also uniform in space. In the following expressions, we are going to employ the following notation for the matrix elements evaluated at the same point of the Brillouin zone,\begin{align}
   \left\langle u_{n_1,\zeta}\left(\mathbf{q}\right)\right|\hat{\mathcal{O} } \left | u_{n_2,\zeta}\left(\mathbf{q}\right) \right\rangle \equiv\mathcal{O}_{\zeta,\mathbf{q}}^{(n_1,n_2)}.
\end{align}
Similarly, all the summations in $\mathbf{q}$ from this point on are restricted to the moir\'e Brillouin zone.

For the layer-resolved current, first-order perturbation theory in the external fields gives (recovering the spin degeneracy)\begin{align}
\label{eq:j_micro}
    j_i^{(\mu)}\left(\omega\right)=\frac{i}{\omega}\sum_{\nu}\chi^{i,j}_{\mu,\nu}\left(\omega\right)E_j^{(\nu)}\left(\omega\right)+\frac{2ie^2}{\omega A}\sum_{n}\sum_{\zeta}\sum_{\mathbf{q}}\mathcal{D}_{\zeta,\mu,\mathbf{q}}^{i,j;(n,n)}f_{\zeta,\mathbf{q}}^{n}\,E_j^{(\mu)}\left(\omega\right),
\end{align}
where we have introduced the retarded (paramagnetic) current-current correlation function given by\begin{align}
    \chi^{i,j}_{\mu,\nu}\left(\omega\right)=\frac{2e^2}{A}\sum_{n_1,n_2}\sum_{\zeta}\sum_{\mathbf{q}}\frac{\left(f_{\zeta,\mathbf{q}}^{n_1}-f_{\zeta,\mathbf{q}}^{n_2}\right)v_{\zeta,\mu,\mathbf{q}}^{i;(n_1,n_2)}v_{\zeta,\nu,\mathbf{q}}^{j;(n_2,n_1)}}{\hbar \omega+i0^+ +\varepsilon_{\zeta,\mathbf{q}}^{n_1}-\varepsilon_{\zeta,\mathbf{q}}^{n_2}}.
\end{align}
$D_3$ and time-reversal symmetries impose the following relations:\begin{subequations}
\begin{align}
&    \chi^{x,x}_{t,t}\left(\omega\right)=\chi^{y,y}_{t,t}\left(\omega\right)=\chi^{x,x}_{b,b}\left(\omega\right)=\chi^{y,y}_{b,b}\left(\omega\right),\\
& \chi^{x,y}_{t,t}\left(\omega\right)=\chi^{y,x}_{t,t}\left(\omega\right)=\chi^{x,y}_{b,b}\left(\omega\right)=\chi^{y,x}_{b,b}\left(\omega\right)=0,\\
& \chi^{x,x}_{t,b}\left(\omega\right)=\chi^{y,y}_{t,b}\left(\omega\right)=\chi^{x,x}_{b,t}\left(\omega\right)=\chi^{y,y}_{b,t}\left(\omega\right),\\
& \chi^{x,y}_{t,b}\left(\omega\right)=-\chi^{y,x}_{t,b}\left(\omega\right)=-\chi^{x,y}_{b,t}\left(\omega\right)=\chi^{y,x}_{b,t}\left(\omega\right).
\end{align}
\end{subequations}
Comparing the result in Eq.~\eqref{eq:j_micro} with the constitutive relations in the main text allows us to write down Kubo formulas for the total and chiral conductivities,\begin{subequations}\begin{align}
&    \sigma_0\left(\omega\right)=\frac{i}{\omega}\sum_{\mu,\nu}\chi_{\mu,\nu}^{x,x}\left(\omega\right)+\frac{2ie^2}{\omega A}\sum_{n}\sum_{\zeta}\sum_{\mathbf{q}}\mathcal{D}_{\zeta,\mathbf{q}}^{x,x;(n,n)}f_{\zeta,\mathbf{q}}^{n},\\
& \sigma_{\textrm{c}}\left(\omega\right)=\frac{i}{\omega}\chi_{t,b}^{x,y}\left(\omega\right).
\end{align}
\end{subequations}

In order to arrive at the final formulas in Eqs.~(6) of the main text, we just need to isolate the interband contributions. The kernel in the correlator can be decomposed in simple fractions as
\begin{align}
  \frac{1}{\hbar \omega+i0^+ +\varepsilon_{\zeta,\mathbf{q}}^{n_1}-\varepsilon_{\zeta,\mathbf{q}}^{n_2}}=\frac{1}{\varepsilon_{\zeta,\mathbf{q}}^{n_1}-\varepsilon_{\zeta,\mathbf{q}}^{n_2}}-\frac{\hbar \omega}{\left(\varepsilon_{\zeta,\mathbf{q}}^{n_1}-\varepsilon_{\zeta,\mathbf{q}}^{n_2}\right)^2}+\frac{\hbar^2 \omega^2}{\left(\varepsilon_{\zeta,\mathbf{q}}^{n_1}-\varepsilon_{\zeta,\mathbf{q}}^{n_2}\right)^2\left(\hbar \omega+i0^+ +\varepsilon_{\zeta,\mathbf{q}}^{n_1}-\varepsilon_{\zeta,\mathbf{q}}^{n_2}\right)}.
\end{align}
The first term (along with the diamagnetic term in the case of the total conductivity) gives rise to the contribution due to intraband electron motion. The integration of the second term cancels, which can be demonstrated using the following identities imposed by time-reversal symmetry:\begin{subequations}
\begin{align}
    & \varepsilon_{\zeta,\mathbf{q}}^{n}=\varepsilon_{-\zeta,-\mathbf{q}}^{n},\\
    & \boldsymbol{v}_{\zeta,\mu,\mathbf{q}}^{(n_1,n_2)}=-\boldsymbol{v}_{-\zeta,\mu,-\mathbf{q}}^{(n_2,n_1)}.
\end{align}
\end{subequations}
The third term leads to the final result in Eqs.~(6), where we have used time-reversal symmetry in order to simplify the sum in valley index.

\subsection{Electromagnetic response}

\subsubsection{Constitutive relations}
Here, we follow Ref.~10 in the main text and rewrite the constitutive relations in the main text in terms of local quantities, as we can assume $z_0\ll c/\omega$ for the relevant frequencies in the experiment. As in the beginning of the previous section, we write the macroscopic electric field as a function of the vertical coordinate, i.e.,  $\mathbf{E}(z,\omega)$. Expanding in series up to first order in $z_0$, we have\begin{align}
\mathbf{E}^{(t,b)}\left(\omega\right)\equiv\mathbf{E}\left(\pm\frac{z_0}{2},\omega\right)\approx \mathbf{E}\left(0,\omega\right)\pm \frac{z_0}{2}\partial_z \mathbf{E}\left(z,\omega\right)|_{z=0}.
\end{align}
To leading order in $z_0$, we thus write\begin{subequations}\begin{align}
& \mathbf{E}\left(\omega\right)\approx \mathbf{E}\left(0,\omega\right),\\
& \mathbf{E}^{\textrm{(cf)}}\left(\omega\right)\approx z_0\,\partial_z \mathbf{E}\left(z,\omega\right)|_{z=0}.
\end{align}\end{subequations}
The macroscopic field is a solution of the Maxwell equations; in particular, from Faraday's law\begin{align}
\mathbf{\hat{z}}\times\partial_z\mathbf{E}(z,\omega)=i\omega\mathbf{B}\left(z,\omega\right).
\end{align}
We can thus relate $\mathbf{E}^{\textrm{(cf)}}$ with the \textit{averaged} magnetic field $\mathbf{B}\left(\omega\right)\equiv \mathbf{B}\left(0,\omega\right)$,\begin{align}
 \mathbf{E}^{\textrm{(cf)}}\left(\omega\right)\approx i\omega z_0\,\mathbf{B}\left(\omega\right)\times\mathbf{\hat{z}}.
\end{align}
In this approximation, we can substitute the bilayer system by a purely 2D system placed at $z=0$, where $\mathbf{j}(\omega)$ expresses the total charge current or, equivalently, the polarization density,\begin{align}
\mathbf{P}\left(\omega\right)=\frac{i}{\omega}\,\mathbf{j}\left(\omega\right),
\end{align}
while the counter-flow current can be used to coarse grain an orbital magnetization density, i.e.,
\begin{align}
\mathbf{M}\left(\omega\right)=\frac{1}{2}\mathbf{z}_0\times\mathbf{j}^{(t)}\left(\omega\right)-\frac{1}{2}\mathbf{z}_0\times\mathbf{j}^{(b)}\left(\omega\right)=\mathbf{z}_0\times\mathbf{j}^{\textrm{(cf)}}\left(\omega\right).
\end{align}
The constitutive relations to leading order in $z_0\omega/c$ are readily found to be \begin{subequations}\label{const}\begin{align}
& \mathbf{P}\left(\omega\right)=\frac{i}{\omega}\sigma_0\left(\omega\right)\mathbf{E}\left(\omega\right)-z_0\,\sigma_{\textrm{c}}\left(\omega\right)\,\mathbf{B}\left(\omega\right),\\
& \mathbf{M}\left(\omega\right)=i\omega z_0^2\,\sigma_{\textrm{cf}}\left(\omega\right)\,\mathbf{B}\left(\omega\right)+\sigma_{\textrm{c}}\left(\omega\right)z_0\,\mathbf{E}\left(\omega\right).
\end{align}
\end{subequations}

This kind of response is allowed by the chiral (purely rotational) symmetry of the moir\'e patterned material, which allows us to identify vectors ($\mathbf{P}$, $\mathbf{E}$) with pseudo-vectors ($\mathbf{M}$, $\mathbf{B}$). Systems resulting from a twist of opposite sign can be envisioned as partners of opposite chirality connected by the inversion/reflection symmetry broken by the relative rotation. In type-I structures, this symmetry is inversion $i$ along the center of the axis connecting a common center of the hexagons. The total and counter-flow currents transform as\begin{subequations}\begin{align}
& i: \,\mathbf{j} \longrightarrow -\mathbf{j},\\
&i:\, \mathbf{j}^{\textrm{(cf)}} \longrightarrow \mathbf{j}^{\textrm{(cf)}}.
\end{align}
\end{subequations}
The same applies to the conjugate forces. In type-II structures, the pertinent symmetry is mirror reflection, $h$, then\begin{subequations}\begin{align}
& h:\,\mathbf{j} \longrightarrow \mathbf{j},\\
& h:\,\mathbf{j}^{\textrm{(cf)}} \longrightarrow -\mathbf{j}^{\textrm{(cf)}}.
\end{align}
\end{subequations}
In both cases, it follows that\begin{subequations}
\label{eq:symmetries_chirality}
\begin{align}
&  i,h:\, \sigma,\sigma_{\textrm{cf}}\longrightarrow \sigma,\sigma_{\textrm{cf}},\\
&i,h:\, \sigma_{\textrm{c}}\longrightarrow -\sigma_{\textrm{c}}.
\end{align}
\end{subequations}
Therefore, chiral partners present opposite CD.

\subsubsection{Fresnel coefficients}
\begin{figure}
\centerline{\includegraphics[width=0.7\linewidth]{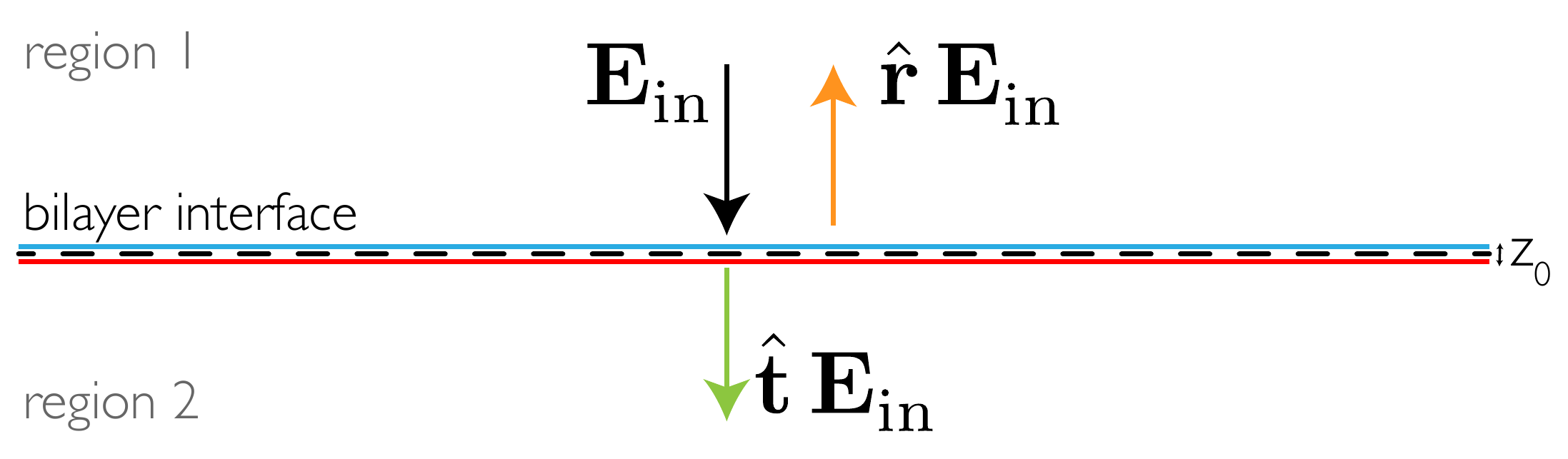}}
\caption{\textbf{Scheme for the derivation of the Fresnel coefficients}. The incident field is partially reflected and transmitted (and absorbed) by the bilayer interface. The bilayer (blue-red) is  understood as a sheet (dashed black) with a coarse-grained chiral response.}
\label{fig:fresnel}
\end{figure}

We now derive the Fresnel coefficients that describe transmission and reflection upon incidence of a normal electromagnetic field on the twisted bilayer, as shown in Fig.~\ref{fig:fresnel}. In the previous section, we have described how the twisted bilayer can be described as a coarse-grained 2D system with a chiral coupling resulting from the Hall drag counter-flow. We employ this insight to model the material system as an interface between two dielectric regions. From now on, we assume that the two regions are vacuum, but it is a trivial extension to consider an embedding medium or substrate.

The reflected and transmitted fields are related to the incident field as
\begin{equation}
\textbf{E}_t=\hat{\textbf{t}}\,\textbf{E}_\text{in}, \quad\quad\textbf{E}_r=\hat{\textbf{r}}\,\textbf{E}_\text{in},
\end{equation}
where the transmission and reflection matrices read
\begin{equation}
\hat{\textbf{t}}=
\begin{bmatrix}
    t       & t_{xy} \\
    -t_{xy}       & t 
\end{bmatrix},\quad
\hat{\textbf{r}}=
\begin{bmatrix}
    r       & r_{xy} \\
    -r_{xy}       & r 
\end{bmatrix},
\end{equation}
and $t,r,r_{xy},t_{xy}$ are the Fresnel coefficients. Note that the structure of the matrices is imposed by $D_3$ and time reversal symmetries. We find the values of the Fresnel coefficients from the constitutive equations \eqref{const} together with the following boundary conditions at the interface:
\begin{subequations}
\begin{align}
&\hat{\textbf{z}}\times(\textbf{H}_2-\textbf{H}_1)=-\ii\omega\textbf{P},\\
&\hat{\textbf{z}}\times(\textbf{E}_2-\textbf{E}_1)=\ii\mu_0\omega\textbf{M},
\end{align}
\end{subequations}
where the subscripts $\{1,2\}$ refer to the regions depicted in Fig.~\ref{fig:fresnel}. The electromagnetic fields in each region are
\begin{subequations}
\begin{align}
&\textbf{E}_1=\textbf{E}_\text{in}+\textbf{E}_r,\quad \textbf{E}_2=\textbf{E}_t,\\
&\mathbf{H}_1=\frac{1}{\mu_0c}\hat{\textbf{z}}\times\left(\textbf{E}_\text{in}-\textbf{E}_r\right),\quad
\mathbf{H}_2=\frac{1}{\mu_0c}\hat{\textbf{z}}\times\textbf{E}_t,
\end{align}
\end{subequations}
where the propagation direction of the incident field is assumed to be normal to the interface. Note that we take both the polarization and magnetization densities to depend only on the incident fields. After some algebra, the Fresnel coefficients are readily found to be
\begin{equation}
r=-\frac{\sigma_0}{2\epsilon_0 c},\quad t=1-\frac{\sigma_0}{2\epsilon_0 c},\quad r_{xy}=0,\quad t_{xy}=\frac{i\omega z_0\sigma_\text{c}}{\epsilon_0 c^2},
\end{equation}
where we have dropped the explicit dependence of the conductivity on the frequency for the sake of clarity.

\subsubsection{Circular dichroism}
The circular dichroism (CD) can be easily derived from the Fresnel coefficients. For a circularly polarized incident field $\textbf{E}_\text{in}^\pm=E_\text{in}\left(\hat{\textbf{x}}\pm i \hat{\textbf{y}}\right)/\sqrt{2}$, the transmitted and reflected fields are readily found by simply operating with the transmission and reflection matrices,
\begin{align}
\textbf{E}_t^\pm=t_\pm\textbf{E}_\text{in}^\pm,\quad\quad\textbf{E}_r^\pm=r_\pm\textbf{E}_\text{in}^\pm,
\end{align}
with
\begin{subequations}
\begin{align}
t_\pm&=t\pm i t_{xy} = 1-\frac{\sigma_0}{2\epsilon_0 c}\mp\frac{\omega z_0}{\epsilon_0 c^2}\sigma_\text{c},\\
r_\pm&=r\pm i r_{xy}=-\frac{\sigma_0}{2\epsilon_0 c}.
\end{align}
\end{subequations}
Note that the reflection coefficient does not depend on the chiral conductivity. This implies that circular dichroism cannot be detected by reflectance-only measurements. One instead has to measure either transmission (i.e., ellipticity) or the absorption of the chiral samples. We focus on the latter to define the circular dichroism as a difference between the absorption of left- and right-circularly polarized light, i.e., 
\begin{align}
\label{eq:ellipticity}
    \text{CD}=\frac{\mathcal{A}_+-\mathcal{A}_-}{(\mathcal{A}_++\mathcal{A}_-)/2},
\end{align}
where $\mathcal{A}_\pm=1-|r_\pm|^2-|t_\pm|^2$ is the absorbance of the bilayer, which reads
\begin{equation}
\mathcal{A}_\pm=\frac{1}{\epsilon_0 c}\text{Re}\left\{\sigma\pm\frac{2\omega z_0}{c}\sigma_\text{c}\mp \frac{\omega z_0}{2\epsilon_0 c^2}\sigma\sigma_\text{c}^*\right\}.
\end{equation}

To first order in $\omega z_0/c$, the circular dichroism is found to be
\begin{align}
    \text{CD}\approx \frac{\omega z_0}{c}\frac{4\,\textrm{Re}\,\sigma_{\textrm{c}}}{\textrm{Re}\,\sigma_{0}}-\frac{\omega z_0}{\epsilon_0 c^2}\frac{\textrm{Re}\,\{\sigma\sigma_{\textrm{c}}^*\}}{\textrm{Re}\,\sigma_{0}}.
\end{align}
Since $|\sigma_0|/\epsilon_0c\ll1$, we drop the second term above and find the expression in the main text.

\subsection{Approximate symmetries and cancellations}

In the context of twisted bilayer graphene, it is common to consider, instead of the full tight-binding dispersion, the linearized theory resulting from the expansion of $\phi_{\mu}$ in Eq.~\eqref{eq:phi} up to first order,\begin{align}
\label{eq:linearization}
    \phi_{(t,b)}\left(\mathbf{p}\right)=\sum_{i}e^{i\boldsymbol{\delta}_i^{(t,b)}\cdot\left(\mathbf{p}+\mathbf{K}_{\zeta}^{(t,b)}\right)}\approx i\sum_{i}\left(\boldsymbol{\delta}_i^{(t,b)}\cdot\mathbf{p}\right)e^{i\boldsymbol{\delta}_i^{(t,b)}\cdot\mathbf{K}_{\zeta}^{(t,b)}}=\frac{\sqrt{3}\,a\,e^{\pm i\zeta\frac{\theta}{2}}}{2}\left(-\zeta p_x+i p_y\right).
\end{align}
Furthermore, it is also common to neglect the extra phase $e^{\pm i\zeta\frac{\theta}{2}}$, which only introduces a subleading correction $\sim\mathcal{O}(\theta^2)$ in the bands. Figure~\ref{fig:bands}~(a) shows the band structures of hBN bilayers for the largest twist angle considered in the main text imposing these approximations. The band dispersion does not change much, particularly in the case of the lowest-energy bands. The agreement improves as the the twist angle decreases.

These approximations have a more subtle effect on the electronic wave functions and the optical response of twisted bilayer graphene due to an electron-hole symmetry. For the following discussion, let us simplify the notation by introducing Pauli matrices $\hat{\mu}_i$ acting on layer degree of freedom such that, for example, layer projectors read just $\hat{P}_t=(\hat{1}+\hat{\mu}_z)/2$, $\hat{P}_{b}=(\hat{1}-\hat{\mu}_z)/2$. In this notation, the electron-hole symmetry, $\{\hat{\mathcal{U}}_{\textrm{e-h}},\hat{\mathcal{H}}_{\zeta}\}=0$, is implemented by the following anti-unitary operator,\begin{align}
  \hat{\mathcal{U}}_{\textrm{e-h}}=i\hat{\mu}_y\otimes\hat{\sigma}_x\,\mathcal{K},
\end{align} 
where $\mathcal{K}$ represents complex conjugation. This approximate symmetry, when enforced, suppresses the optical transitions between bands of opposite energies at the $\boldsymbol{m}$ point and, in combination with C$_{2}$ symmetries, can be used to prove that contributions from opposite points of the moir\'e Brillouin zone to the interband chiral conductivity cancel out. 

\begin{figure}
\centerline{\includegraphics[width=\linewidth]{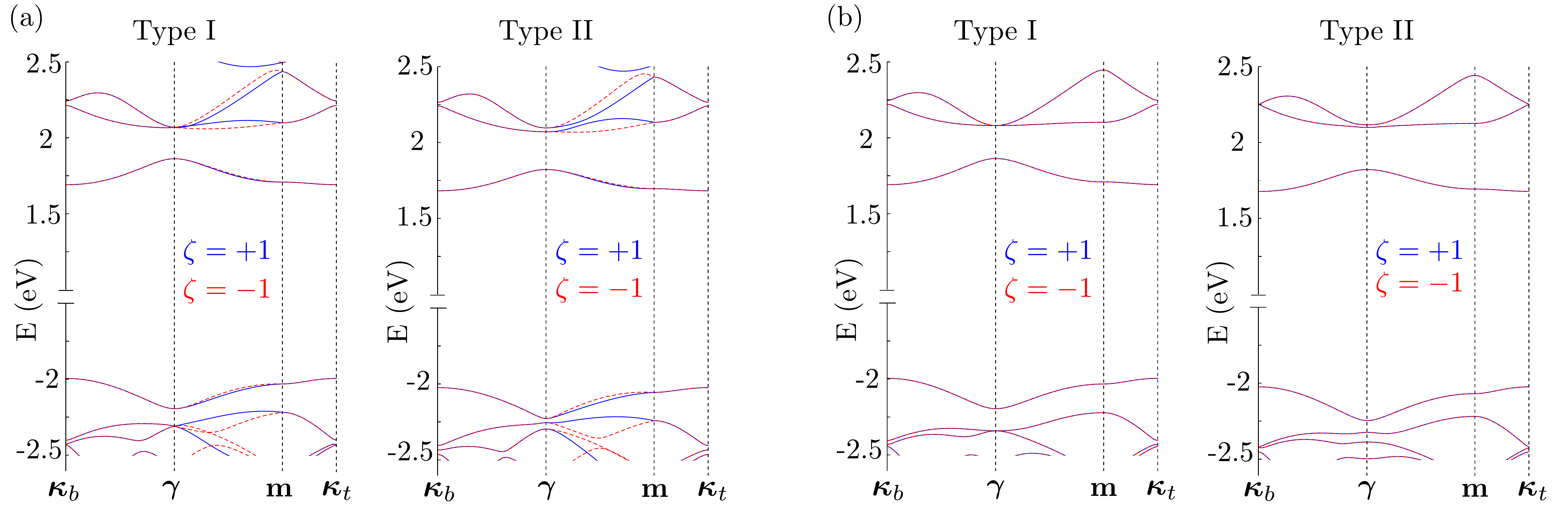}}
\caption{\textbf{Bands of twisted hBN at $\theta=7.34^{\textrm{o}}$ in a simplified model.} (a) Band structures obtained by taking Eq.~\eqref{eq:linearization}, neglecting also the twist-angle related phase. (b) The same by taking now $w_{\textrm{AA'}}=0$ (type I) and $w_{\textrm{AB}}=0$ (type II).}
\label{fig:bands}
\end{figure}

As mentioned in the main text, the electron-hole symmetry is not present in twisted hBN due to the physical inequivalence of the two sublattices. In particular, in the case of type-I moir\'e structures, this is just due to the opposite sign of $\Delta(\mathbf{r})$ in the top and bottom layers; in the case of type-II structures, due to the fact that $w_{\textrm{BB}}\neq w_{\textrm{NN}}$. However, as we discuss next, there are approximate \textit{pseudo-inversion} symmetries in certain limits of the model, which give rise to cancellations of the chiral response and, in the case of type-II structures, to additional selection rules that suppress optical transitions at certain high-symmetry points of the moir\'e Brillouin zone.

\subsubsection{Type I}

In the limit of small twist angles, the Hamiltonian of type-I structures can be simplified to\begin{align}
\label{eq:typeI_simplification}
    \hat{\mathcal{H}}_{\zeta}\left(\mathbf{r}\right)=\frac{i\sqrt{3}a t}{2}\,\boldsymbol{\hat{\sigma}}_{\zeta}\cdot\boldsymbol{\partial}+\frac{\Delta\left(\mathbf{r}\right)}{2}\,\hat{\mu}_z\otimes\hat{\sigma}_z+\frac{1}{2}\hat{\mu}_x\otimes\left(\hat{T}_{\zeta}\left(\mathbf{r}\right)+\hat{T}_{\zeta}^{\dagger}\left(\mathbf{r}\right)\right)+\frac{i}{2}\hat{\mu}_y\otimes\left(\hat{T}_{\zeta}\left(\mathbf{r}\right)-\hat{T}_{\zeta}^{\dagger}\left(\mathbf{r}\right)\right),
\end{align}
where $\boldsymbol{\hat{\sigma}}_{\zeta}=(\zeta\hat{\sigma}_x,\hat{\sigma}_y)$. The bands shown in Fig.~\ref{fig:bands}~(a) (left panel) are deduced from this Hamiltonian.

If we take now $w_{\textrm{AA'}}=0$, the band dispersion of the lowest-energy bands do not change much, but as shown in the calculation of Fig.~\ref{fig:bands}~(b) (left panel), the spectrum becomes degenerate in valley over the whole moir\'e Brillouin zone. This is due to a new \textit{pseudo-inversion} symmetry that relates the Hamiltonians of the two valley sectors. Specifically, if $w_{\textrm{AA'}}=0$, then the Hamiltonian in Eq.~\eqref{eq:typeI_simplification} satisfies\begin{subequations}\begin{align}
& \hat{\mathcal{U}}\,\hat{\mathcal{H}}_{\zeta}\left(\mathbf{r}\right)\hat{\mathcal{U}}^{\dagger}=\hat{\mathcal{H}}_{-\zeta}\left(\mathbf{r}\right),\\
& \text{with}\,\,\,\hat{\mathcal{U}}=\hat{\mu}_y\otimes\hat{\sigma}_y.
\end{align}
\end{subequations}
When the symmetry is enforced, the chiral response is suppressed. This result follows from the following property of the velocity operators:\begin{align}
    \hat{\mathcal{U}}\,\boldsymbol{\hat{v}}_{\zeta,t}\hat{\mathcal{U}}^{\dagger}=\boldsymbol{\hat{v}}_{-\zeta,b},
\end{align}
which can be used, along with C$_2$ symmetries, to prove that the contributions to $\chi_{t,b}^{x,y}(\omega)$ from the two valleys cancel each other. 

\subsubsection{Type II}

In the case of type-II structures, the Hamiltonian can be simplified to\begin{align}
\label{eq:typeII_simplification}
    \hat{\mathcal{H}}_{\zeta}\left(\mathbf{r}\right)=\frac{i\sqrt{3}a t}{2}\,\boldsymbol{\hat{\sigma}}_{\zeta}\cdot\boldsymbol{\partial}+\frac{\Delta\left(\mathbf{r}\right)}{2}\,\hat{\sigma}_z+\frac{1}{2}\hat{\mu}_x\otimes\left(\hat{T}_{\zeta}\left(\mathbf{r}\right)+\hat{T}_{\zeta}^{\dagger}\left(\mathbf{r}\right)\right)+\frac{i}{2}\hat{\mu}_y\otimes\left(\hat{T}_{\zeta}\left(\mathbf{r}\right)-\hat{T}_{\zeta}^{\dagger}\left(\mathbf{r}\right)\right),
\end{align}
The bands in Fig.~\ref{fig:bands}~(a) (right panel) are deduced from this Hamiltonian.

If we take now $w_{\textrm{AB}}=0$, just as in the previous case the dispersion of the lowest-energy bands does not change significantly, but the spectrum becomes also valley degenerate, see Fig.~\ref{fig:bands}~(b) (right panel). Again, this is due to a new \textit{pseudo-inversion} symmetry. Specifically, if $w_{\textrm{AB}}=0$, then the Hamiltonian in Eq.~\eqref{eq:typeII_simplification} satisfies\begin{subequations}\begin{align}
& \hat{\mathcal{U}}\,\hat{\mathcal{H}}_{\zeta}\left(\mathbf{r}\right)\hat{\mathcal{U}}^{\dagger}=\hat{\mathcal{H}}_{\zeta}\left(-\mathbf{r}\right),\\
& \text{with}\,\,\,\hat{\mathcal{U}}=\hat{\mu}_x\otimes\hat{\sigma}_z.
\end{align}
\end{subequations}
Note that, contrary to the previous case, the symmetry is nonlocal in space but imposes relations between the eigenvalues and eigenstates of the same valley. In particular, it implies that $\varepsilon_{\zeta,\mathbf{q}}^{n}=\varepsilon_{\zeta,-\mathbf{q}}^{n}$. This, along with time-reversal symmetry, implies that $\varepsilon_{\zeta,\mathbf{q}}^{n}=\varepsilon_{-\zeta,\mathbf{q}}^{n}$.

When the symmetry is enforced, the chiral response is also suppressed. This result follows from the following property of the velocity operators:\begin{align}
\label{eq:property}
    \hat{\mathcal{U}}\,\boldsymbol{\hat{v}}_{\zeta,t}\hat{\mathcal{U}}^{\dagger}=-\boldsymbol{\hat{v}}_{\zeta,b},
\end{align}
which can be used to prove that the contributions to $\chi_{t,b}^{x,y}(\omega)$ from opposite $\mathbf{q}$ points within the same valley cancel to each other.

The relation in Eq.~\eqref{eq:property} can also be used to demonstrate that some of the optical transitions at $\boldsymbol{\gamma}$ and $\boldsymbol{m}$ points (which remain invariant under the symmetry operation) are forbidden when the symmetry is enforced. In particular, since the lowest conduction and valence band states have the same parity under this symmetry, namely,\begin{align}
    & \hat{\mathcal{U}}\left|u_{n,\zeta}\left(\mathbf{q}\right)\right\rangle=-\left|u_{n,\zeta}\left(\mathbf{q}\right)\right\rangle\,\,\, \text{for}\,\, \mathbf{q}=\boldsymbol{\gamma},\boldsymbol{m},\,\,\, \text{and}\,\, n=c,v,
\end{align}
it can be easily seen that $\boldsymbol{v}_{\zeta,\mathbf{q}}^{(c,v)}=0$ for $\mathbf{q}=\boldsymbol{\gamma},\boldsymbol{m}$. This gives rise to a partial suppression of the total conductivity of type-II structures (more evident as the twist angle decreases), which contribute to enhance the circular dichroism with respect to type-I structures.

\end{document}